\documentclass[11pt]{article}

\usepackage{amsthm}
\usepackage{mathtools}

\usepackage[T1]{fontenc}
\usepackage[utf8]{inputenc}
\usepackage{newtxtext, newtxmath}
\usepackage[protrusion=true, expansion=true]{microtype}
\usepackage{setspace}

\usepackage[english]{babel}
\usepackage{csquotes}

\usepackage[margin=1in]{geometry}

\usepackage{booktabs}
\usepackage[hang, small, labelfont=bf, textfont=it, up]{caption}
\usepackage{enumitem}
\setlist[itemize]{noitemsep}
\usepackage{graphicx}
\usepackage{longtable}
\usepackage{ragged2e}
\usepackage{rotating}
\usepackage{threeparttable}

\usepackage{abstract}

\usepackage{titlesec}
\titleformat*{\section}{\large\bfseries}
\titleformat*{\subsection}{\bfseries}

\usepackage[usenames, dvipsnames]{xcolor}

\usepackage[hidelinks]{hyperref}

\usepackage[authordate, backend=biber]{biblatex-chicago}
\usepackage{titling}

\title{
\Large\bfseries
Latent Unbalancedness in Three-Way Gravity Models
\thanks{We thank Thomas K. Bauer and Joschka Wanner for helpful comments.}
}
\author{
\normalsize
Daniel Czarnowske
\thanks{
Heinrich-Heine-Universität Düsseldorf, Universitätsstr. 1, 40225 Düsseldorf, Germany, phone: +49 211 81-10620, e-mail: \texttt{\href{mailto:daniel.czarnowske@hhu.de}{daniel.czarnowske@hhu.de}}
}
\and
Amrei Stammann
\thanks{
Ruhr-Universität Bochum, Universitätsstraße 150, 44801 Bochum, Germany, phone: +49 234 32-22875, e-mail: \texttt{\href{mailto:amrei.stammann@rub.de}{amrei.stammann@rub.de}}
}
}
\date{\small\today}

\DeclareMathOperator{\iid}{\text{iid.}\;}

\DeclareMathOperator{\EX}{\mathbb{E}}

\DeclareMathOperator{\N}{\mathcal{N}}

\theoremstyle{definition}
\newtheorem*{algorithm}{Algorithm}

\begin{document}

\maketitle
\thispagestyle{empty}

\renewcommand{\abstractname}{\vspace{-5em}}
\begin{abstract}
\small
\noindent Many panel data sets used for pseudo-poisson estimation of three-way gravity models are implicitly unbalanced because uninformative observations  are redundant for the estimation. We show with real data as well as simulations that this phenomenon, which we call latent unbalancedness, amplifies the inference problem recently studied by \textcite{wz2021}.\\[0.5em]
\footnotesize
\vfill
\end{abstract}

\clearpage
\onehalfspacing
\setcounter{page}{1}

\section{Introduction}
\label{sec:introduction}

Consider the ``three-way'' gravity model 
\begin{equation}
    \label{eq:estimation_equation_gravity}
    y_{ijt} \coloneqq \exp(\beta \, x_{ijt} + \alpha_{it} + \gamma_{jt} + \eta_{ij}) \, \omega_{ijt} \, , \; i, j \in \{1, \ldots, N\} \, , \; t \in \{1, \ldots, T\} \, ,
\end{equation}
where $y_{ijt}$ is the trade flow from country $i$ (exporter) to country $j$ (importer) at time $t$, $x_{ijt}$  a set of trade cost variables, $\beta$  the corresponding trade cost parameters, $\alpha_{it}$, $\gamma_{jt}$, and $\eta_{ij}$ the three sets of fixed effects capturing market sizes, multilateral resistances, and other (unobserved) time-invariant trade costs, respectively, and $\omega_{ijt}$  a multiplicative idiosyncratic error term.\footnote{A theoretical motivation of \eqref{eq:estimation_equation_gravity} is provided in \textcite{hm2014}.} \textcite{wz2021} (hereafter WZ) study the pseudo-poisson maximum likelihood estimator (hereafter FE-PPML) for the model parameters in \eqref{eq:estimation_equation_gravity}.\footnote{FE-PPML is an estimator in spirit of \textcite{gmt1984}, thus only the conditional mean $\EX[y_{ijt} \, | \, x_{ijt}, \alpha_{it}, \gamma_{jt}, \eta_{ij}] = \exp(\beta \, x_{ijt} + \alpha_{it} + \gamma_{jt} + \eta_{ij})$ has to be correctly specified.}  Under asymptotics where $N \rightarrow \infty$ and $T$ is fixed, WZ show that the estimator of $\beta$, $\hat{\beta}$, is consistent but that its distribution is distorted by a constant bias term of order $1 / N$. To improve inference, they suggest to debias $\hat{\beta}$ using a bias correction. Their simulations show, that especially for samples with small $N$, inference can be significantly improved by debiasing $\hat{\beta}$.\footnote{For completeness, WZ also show a downward bias in cluster-robust covariance estimators, which further affects inference.}

One key advantage of FE-PPML is its genuine way of dealing with zero trade flows \parencite{st2006}, which is an important feature of the estimator, because in practice  at least some countries do not trade with each other. However, zero trade flows can  lead to non-unique infinite FE-PPML estimates for $\alpha_{it}$, $\gamma_{jt}$, and $\eta_{ij}$. Consequently, all observations that are affected by these nonexistent estimates have zero log-likelihood contribution and thus are uninformative for the estimation of $\beta$.\footnote{This is essentially the complete separation case described in \textcite{aa1984} and \textcite{st2010} for binary-choice and poisson models, respectively.} Since this phenomenon is not directly apparent to the researcher, we refer to it as \textit{latent unbalancedness} to distinguish it from \textit{classical unbalancedness}, where observations are missing or incomplete.\footnote{WZ (Section 3.4.4) briefly note that their bias corrections can also be applied to panel data with missing values, but do not discuss the consequences for inference.}


The remainder of this paper is as follows. Section \ref{sec:inference} explains latent unbalancedness and its consequences for inference. Section \ref{sec:simulation_evidence} demonstrates the inference problem  in a small-scale simulation study.

\section{Latent Unbalancedness and its Consequences for Inference}
\label{sec:inference}

\subsection{Latent Unbalancedness}

Uninformative observations and hence latent unbalancedness are a direct consequence of zero trade flows, which can be derived directly from the first-order conditions (FOC's) of the FE-PPML:
\begin{align}
    \label{eq:feppml_focs}
    \hat{\beta} \colon& \sum_{i, j}^{N} \sum_{t = 1}^{T} (y_{ijt} - \lambda_{ijt}) x_{ijt} = 0 \, ,&
    \hat{\alpha}_{it} \colon& \sum_{j = 1}^{N} y_{ijt} - \lambda_{ijt} = 0 \, ,\\
    \hat{\gamma}_{jt} \colon& \sum_{i = 1}^{N} y_{ijt} - \lambda_{ijt} = 0 \, ,& \hat{\eta}_{ij} \colon& \sum_{t = 1}^{T} y_{ijt} - \lambda_{ijt} = 0 \, , \nonumber
\end{align}
where $\lambda_{ijt} \coloneqq \exp(\beta \, x_{ijt} + \alpha_{it} + \gamma_{jt} + \eta_{ij})$. From equation \eqref{eq:feppml_focs} we can  infer that,  if a country pair $ij$ never trades at any point in time, than the FOC of $\hat{\eta}_{ij}$ is trivially satisfied if we set $\eta_{ij} = - \infty$, i.e. the corresponding estimate does not exist. Consequently, all other parameters of $\{\lambda_{ijt} \colon t \in \{1, \ldots, T\}\}$ become irrelevant for the estimation, since $\eta_{ij} = - \infty$ is sufficient to predict all zero trade flows independently of the remaining model parameters. Thus, for the estimation of $\beta$, all affected observations are uninformative and may as well be discarded from the sample. Similar arguments can be used for the FOC's of $\hat{\alpha}_{it}$ and $\hat{\gamma}_{jt}$. For instance, if a country $i$ does not export to any country at time $t$, we can set $\alpha_{it} = - \infty$, and $\gamma_{jt} = - \infty$ if a country $j$ does not import from any other country at time $t$. 

Furthermore, it is likely that after excluding uninformative observations, some country pairs appear only for one time period, i.e. become \textit{singletons}. Independent of the remaining model parameters, $\eta_{ij}$ alone is sufficient to predict the corresponding trade flow for these singletons. Hence, these  observations have zero log-likelihood contribution and thus are uninformative for the estimation of $\beta$. Again, we can use similar arguments for an exporting country $i$ at time $t$ that exports only to one country, or an importing country $j$ at time $t$ that only imports from one country. The supplement sketches a simple iterative algorithm that we used to detect uninformative observations.


\subsection{Consequences for Inference}

Although latent unbalancedness might not be a problem from an estimation point of view, it has consequences for inference.\footnote{$\hat{\beta}$ is identical up to some numerical precision, whether uninformative observations are excluded or not. \textcite{st2010} note that, uninformative observations can cause convergence problems. Therefore,  they recommend to exclude these observations.} Uninformative observations  have zero log-likelihood contribution and thus do not fulfill the assumption that the corresponding second-order derivatives are bounded away from zero (see Theorem 1 / Assumption (iii) in the Appendix of WZ). 

To show what latent unbalancedness implies for WZ's asymptotic bias result, it is useful to introduce their heuristic. Let $n$, $p_{\alpha}$, and $p_{\gamma}$ denote the sample size, the number of exporter-time fixed effects, and the number of importer-time fixed, respectively. The order of the bias can be approximated by $\text{bias}(\hat{\beta}) \sim p_{\alpha} / n + p_{\gamma} / n$. Applied to a balanced panel with $n = N^{2}T$, $p_{\alpha} = NT$, and $p_{\gamma} = NT$, we get $\text{bias}(\hat{\beta}) \sim 1 / N + 1 / N$, which is exactly the order of the asymptotic bias derived by WZ. It appears further useful to distinguish between the number of exporters ($I = N$) and importers ($J = N$). Let $n^{\ast} < IJT$, $p_{\alpha}^{\ast} < IT$, and $p_{\gamma}^{\ast} < JT$ denote the sample size, the number of exporter-time fixed effects, and the number of importer-time fixed effects after excluding uninformative observations, respectively. With $n = n^{\ast}$, $p_{\alpha} = p_{\alpha}^{\ast}$, and $p_{\gamma} = p_{\gamma}^{\ast}$, we then get $\text{bias}(\hat{\beta}) \sim 1 / \overline{J} + 1 / \overline{I}$, where $\overline{I} \coloneqq n^{\ast} / p_{\gamma}^{\ast} < N$ and $\overline{J} \coloneqq n^{\ast} / p_{\alpha}^{\ast} < N$ are the average number of exporters and importers, respectively.\footnote{A simulation study in the supplement supports the validity of the heuristic even for unbalanced panels.} The leading order of the bias is given by $1 / \min(\overline{I}, \overline{J})$. Thus, depending on how many observations are uninformative, the bias can be substantially larger than would be expected given the size of the initially balanced sample. As a consequence, researchers may incorrectly conclude that the asymptotic bias in their analysis is negligible, simply because their initial sample has a large $N$.

\begin{figure}[!htbp]
\caption{Consequences of Latent Unbalancedness in Industry-level Trade Data}
\centering
\includegraphics[width=0.9\textwidth]{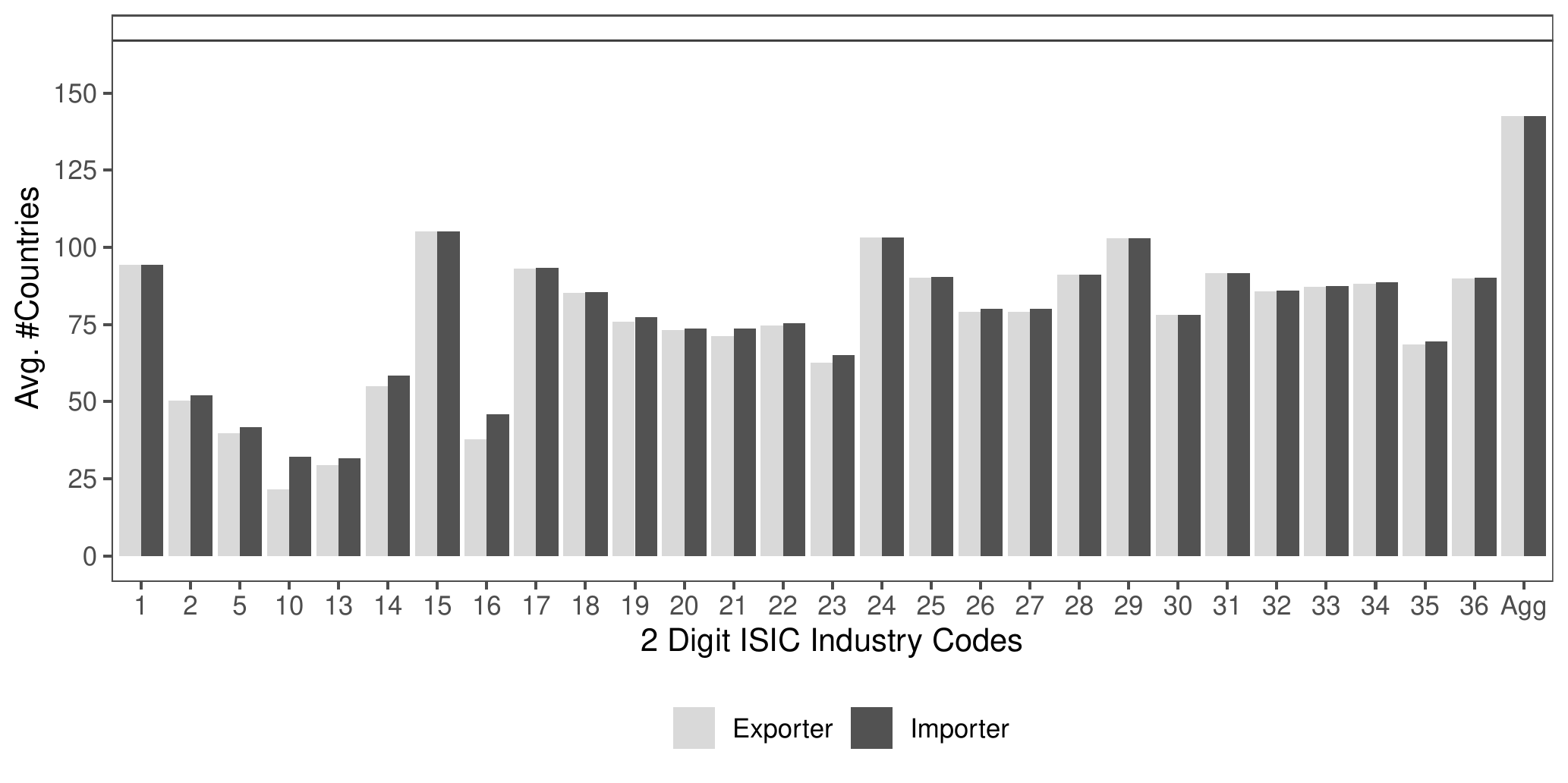}
\begin{minipage}{0.9\textwidth}
	\footnotesize
	\textbf{Notes:} Data taken from the replication package of WZ. Figure shows average number of exporters ($\overline{I}$) and importers ($\overline{J}$) for various 2 digit \textit{ISIC} industries plus aggregate trade after excluding uninformative observations. The solid line indicates $N = 167$, i.e. the initial number of trading countries in the data.
\end{minipage}
\label{fig:latent_unbalancedness}
\end{figure}

Figure \ref{fig:latent_unbalancedness} shows $\overline{I}$ and $\overline{J}$ in the industry-level trade data used by WZ, after excluding uninformative observations. Note that each sample initially covers $N = 167$ countries that trade with each other for $T = 5$ points in time. We find that dis-aggregated industry-level data is heavily affected by latent unbalancedness. This is not surprising, because the more trade flows are dis-aggregated, the more likely it is that some countries either do not have a particular industry or that some countries do not import certain products. On average across all industries, the order of the leading bias is $1 / 75$. This is more than twice than $1 / 167$, which is the order of the bias we would expect from the initial sample. The consequences for inference are most severe in the coal industry (industry 10). Here, the order of the leading bias is $1 / 22$, which is almost eight times $1 / 167$. The inference problem is least pronounced for the aggregated trade data with a leading bias of $1 / 143$. 

\section{Simulation Evidence}
\label{sec:simulation_evidence}

A simulation study demonstrates the inference problem. To mimic latent unbalancedness, we first generate balanced panels using DGP I of WZ,
\begin{align}
    y_{ijt} &= \lambda_{ijt} \, \omega_{ijt} \, , \\
    x_{ijt} &= 0.5 \, x_{ijt-1} + \alpha_{it} + \gamma_{jt} + \eta_{ij} + \nu_{ijt} \, , \nonumber \\
    \omega_{ijt} &= \exp(- 0.5 \, \log(1 + \lambda_{ijt}^{- 2}) + (\log(1 + \lambda_{ijt}^{- 2}))^{- 0.5} \, z_{ijt}) \, , \nonumber \\
    z_{ijt} &= 0.3 \, z_{ijt-1} + \xi_{ijt} \, , \nonumber
\end{align}
where $i, j \in \{1, \ldots, N\}$, $t \in \{1, \ldots, T\}$, $\lambda_{ijt} = \exp(\beta \, x_{ijt} + \alpha_{it} + \gamma_{jt} + \eta_{ij})$, $\alpha_{it}$, $\gamma_{jt}$, and $\eta_{ij}$ are $\sim \iid \N(0, 0.25)$, $\nu_{ijt} \sim \iid \N(0, \sqrt{0.5})$, $x_{ij0} = \eta_{ij} + \nu_{ij0}$, $z_{ij0} \sim \iid \N(0, 1)$, $\xi_{ijt} \sim \iid \N(0, 0.91)$, and $\N(a, b)$ denotes the normal distribution with mean $a$ and variance $b$. Then, we set all trade flows of $\psi N^2$ country pairs with the smallest values of $\eta_{ij}$ to zero. Consequently, these observations become uninformative and can be excluded so that the final sample has $n^{\ast} = (1 - \psi) N^2 T \approx N \overline{N} T$ observations.\footnote{This is essentially an attrition process where uninformative observations are conditionally missing at random.} We consider different fractions of missing observations $\psi \in \{0, 0.1, \ldots, 0.8, 0.9\}$ and set $N = 200$, $T = 5$, and $\beta = 1$. Thus, $\psi = 0.9$ is comparable to the situation in the coal industry, where about 90\% of the observations are uninformative. We analyze relative biases and ratios of bias to standard deviation for same estimators as WZ: an uncorrected estimator (FE-PPML), an analytically bias-corrected estimator (ABC), and a split-panel jackknife bias-corrected estimator (SPJ). Our results are based on $10{,}000$ simulated samples for each $\psi$.\footnote{More detailed results are reported in the supplement.}

\begin{figure}[!htbp]
\caption{Relative Bias (in \%) and Bias / SD of $\hat{\beta}$ for Different $\psi$}
\centering
\includegraphics[width=0.9\textwidth]{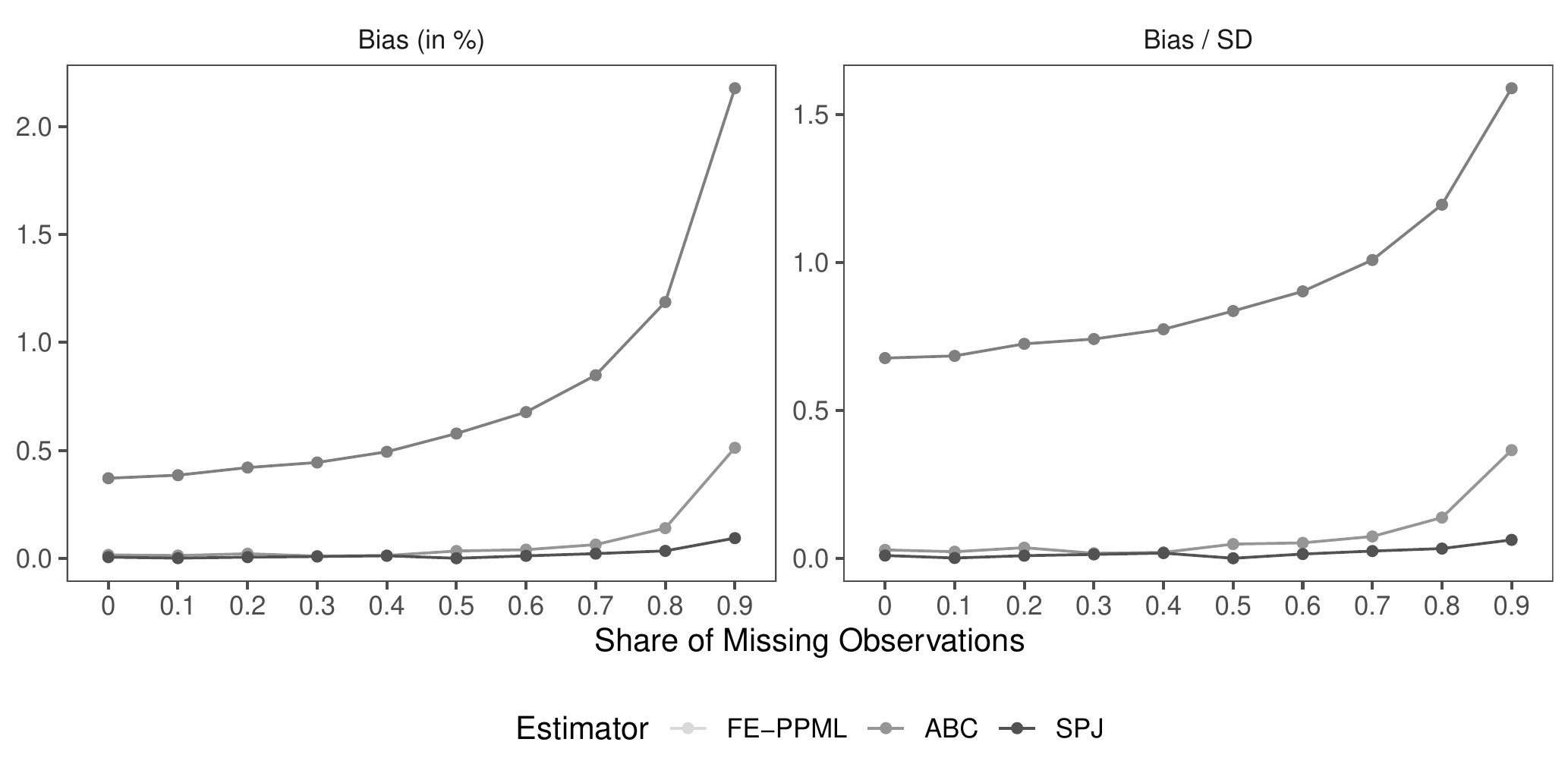}
\begin{minipage}{0.9\textwidth}
	\footnotesize
	\textbf{Notes:} Figure shows relative biases (in \%) and ratios of bias to standard deviation for different estimators for $\beta$: uncorrected FE-PPML, analytically corrected ABC, and split-panel jackknife corrected SPJ. $N = 200$, $T = 5$, and $\psi \in \{0, 0.1, \ldots, 0.8, 0.9\}$. Results are based on $10{,}000$ simulated samples for each $\psi$.
\end{minipage}
\label{fig:results}
\end{figure}

Figure \ref{fig:results} summarizes the results. We find that biases increase with $\psi$, which is to be expected, given the heuristic, as the average number of trading countries ($\overline{N}$) decreases. Furthermore, the biases also increase relative to the dispersion of the estimators, which means that the centering of the standardized estimator, and thus the inference, gets worse as $\psi$ increases. The results demonstrate the importance of identifying latent unbalancedness in empirical work.

\section{Concluding Remarks}
\label{sec:conclusion}


We pointed out the relevance of latent unbalancedness for pseudo-poisson estimation of three-way gravity models. Researchers who are unaware of latent unbalancedness and thus rely on the initial sample size to assess the severity of the inference problem - recently shown by WZ - may be misled and mistakenly decide against using bias-corrected estimators. Latent unbalancedness appears to be especially prevalent in sparse networks, e.g. when analysing highly dis-aggregated trade flows as in the empirical application of WZ. Moreover, latent unbalancedness is clearly not only a problem for three-way FE-PPML, but also for other fixed effects maximum likelihood estimators. Finally, we should emphasize that our findings also carry over to classical unbalancedness, i.e. unbalancedness due to missing or incomplete observations, and that uninformative observations can also arise from other more complicated separation issues, as recently shown by \textcite{cgz2021}.

\clearpage
\printbibliography

\clearpage
\appendix
\begin{center}
    \textbf{\LARGE Online Supplement for ``Latent Unbalancedness in Three-Way Gravity Models''}\\[1em]
    {\Large by Daniel Czarnowske and Amrei Stammann}
\end{center}
\vspace{5em}

\section{Iterative Algorithm to Detect Uninformative Observations}

We present a simple algorithm to detect all uninformative observations based on the definition in Section 2.2. The idea is as follows. We start with the initial sample and then work sequentially through the FOC's of $\hat{\alpha}_{it}$, $\hat{\gamma}_{jt}$, and $\hat{\eta}_{ij}$. For instance, we compute $s_{it} = \sum_{j} y_{ijt}$ for each exporter-time combination $it$. If $s_{it} > 0$, we keep the corresponding exporter-time observations, otherwise ($s_{it} = 0$) we drop them. After we worked through all FOC's, we check for singleton observations and drop them as well. We repeat these steps until we do not find any additional uninformative observation.  

\begin{algorithm}
(Detecting Uninformative Observations)
\begin{itemize}[itemindent=2em]
    \item[Step 0:] Set $r = 0$ and $n^{r}$ to the initial sample size, i.e. $n^{r} = 138{,}610$ in our example. Repeat step 1--5 until $n^{r} = n^{r - 1}$, i.e. we do not find any additional uninformative observations.
    \item[Step 1:] For each exporter-time combination $it$, compute the sum of all trade flows $s_{it}$. If $s_{it} = 0$, drop all observations of the exporter-time combination $it$, otherwise keep them.
    \item[Step 2:] For each importer-time combination $jt$, compute the sum of all trade flows $s_{jt}$. If $s_{jt} = 0$, drop all observations of the importer-time combination $jt$, otherwise keep them.
    \item[Step 3:] For each country pair $ij$, compute the sum of all trade flows $s_{ij}$. If $s_{ij} = 0$, drop all observations of country pair $ij$, otherwise keep them.
    \item[Step 4:] Check all exporter-time combinations, importer-time combinations, and country pairs for singleton observations and drop them.
    \item[Step 5:] Set $r = r + 1$ and $n^{r}$ to the current sample size.
\end{itemize}
\end{algorithm}

\clearpage

\section{Latent Unbalancedness in Industry-Level Trade Data}

\begin{table}[!htbp]
\caption{Sample Sizes and Average Number of Exporters and Importers}
\centering
\begin{threeparttable}
\begin{tabular}{lccccc}
  \toprule
Industry & Code & $n^{\ast}$ & $n^{\ast} / {n}$ & $\overline{I}$ & $\overline{J}$ \\ 
  \midrule
  Agriculture &   1  & 78{,}765 & 0.568  & 94 & 94 \\ 
  Forestry &   2  & 41{,}707 & 0.301  & 50 & 52 \\ 
  Fishing &   5 & 32{,}313 & 0.233 & 40 & 42 \\ 
  Coal &  10 & 15{,}892 & 0.115  & 22 & 32 \\ 
  Metal Ores &  13 & 20{,}357 & 0.147 & 29 & 32 \\ 
  Other Mining \& Quarrying &  14  & 45{,}933 & 0.331  & 55 & 58 \\ 
  Food \& Beverages &  15 & 87{,}915 & 0.634  & 105 & 105 \\ 
  Tobacco &  16  & 31{,}256 & 0.226  & 38 & 46 \\ 
  Textiles &  17  & 77{,}659 & 0.560 & 93 & 93 \\ 
  Apparel &  18 & 71{,}111 & 0.513 & 85 & 86 \\ 
  Leather Products &  19  & 63{,}452 & 0.458  & 76 & 77 \\ 
  Wood \& Cork Products &  20  & 61{,}175 & 0.441  & 73 & 74 \\ 
  Paper \& Paper Products &  21  & 59{,}554 & 0.430 & 71 & 74 \\ 
  Printed \& Recorded Media &  22  & 62{,}318 & 0.450  & 75 & 75 \\ 
  Coke \& Refined Petroleum &  23  & 52{,}214 & 0.377 & 63 & 65 \\ 
  Chemicals \& Chemical Products &  24  & 86{,}185 & 0.622  & 103 & 103 \\ 
  Rubber \& Plastic Products &  25  & 75{,}311 & 0.543  & 90 & 91 \\ 
  Non-Metallic Mineral Products &  26 & 66{,}136 & 0.477  & 79 & 80 \\ 
  Basic Metal Products &  27 & 66{,}078 & 0.477  & 79 & 80 \\ 
  Fabricated Metal Products (excl. Machinery) &  28  & 76{,}095 & 0.549 & 91 & 91 \\ 
  Machinery \& Equipment n. e. c. &  29 & 85{,}955 & 0.620  & 103 & 103 \\ 
  Office, Accounting \& Computer Equipment &  30 & 65{,}219 & 0.471 & 78 & 78 \\ 
  Electrical Equipment &  31 & 76{,}485 & 0.552 & 92 & 92 \\ 
  Communications Equipment &  32 & 71{,}600 & 0.517 & 86 & 86 \\ 
  Medical \& Scientific Equipment &  33 & 72{,}790 & 0.525 & 87 & 87 \\ 
  Motor Vehicles, Trailers \& Semi-trailers &  34 & 73{,}715 & 0.532 & 88 & 89 \\ 
  Other Transport Equipment &  35 & 57{,}217 & 0.413 & 69 & 70 \\ 
  Furniture \& Other Manufacturing &  36 & 75{,}156 & 0.542 & 90 & 90 \\ 
  Aggregate & Agg  & 119{,}040 & 0.859 & 143 & 143 \\ 
  \bottomrule
\end{tabular}
\begin{tablenotes}
\footnotesize
\item \textbf{Notes:} Data taken from the replication package of WZ. Each data set has $n = 138{,}610$ observations and includes all bilateral trade flows of $N = 167$ countries during years 1995, 2000, 2005, 2010, and 2015 ($T = 5$). $n^{\ast}$, $\overline{I}$, and $\overline{J}$ denote the sample size, the average number of exporters and average number of importers after excluding uninformative observations, respectively.
\end{tablenotes}
\end{threeparttable}
\end{table}

\begin{figure}[!htbp]
\caption{Latent Unbalancedness in Industries 10, 13, and 16}
\centering
\includegraphics[width=0.9\textwidth]{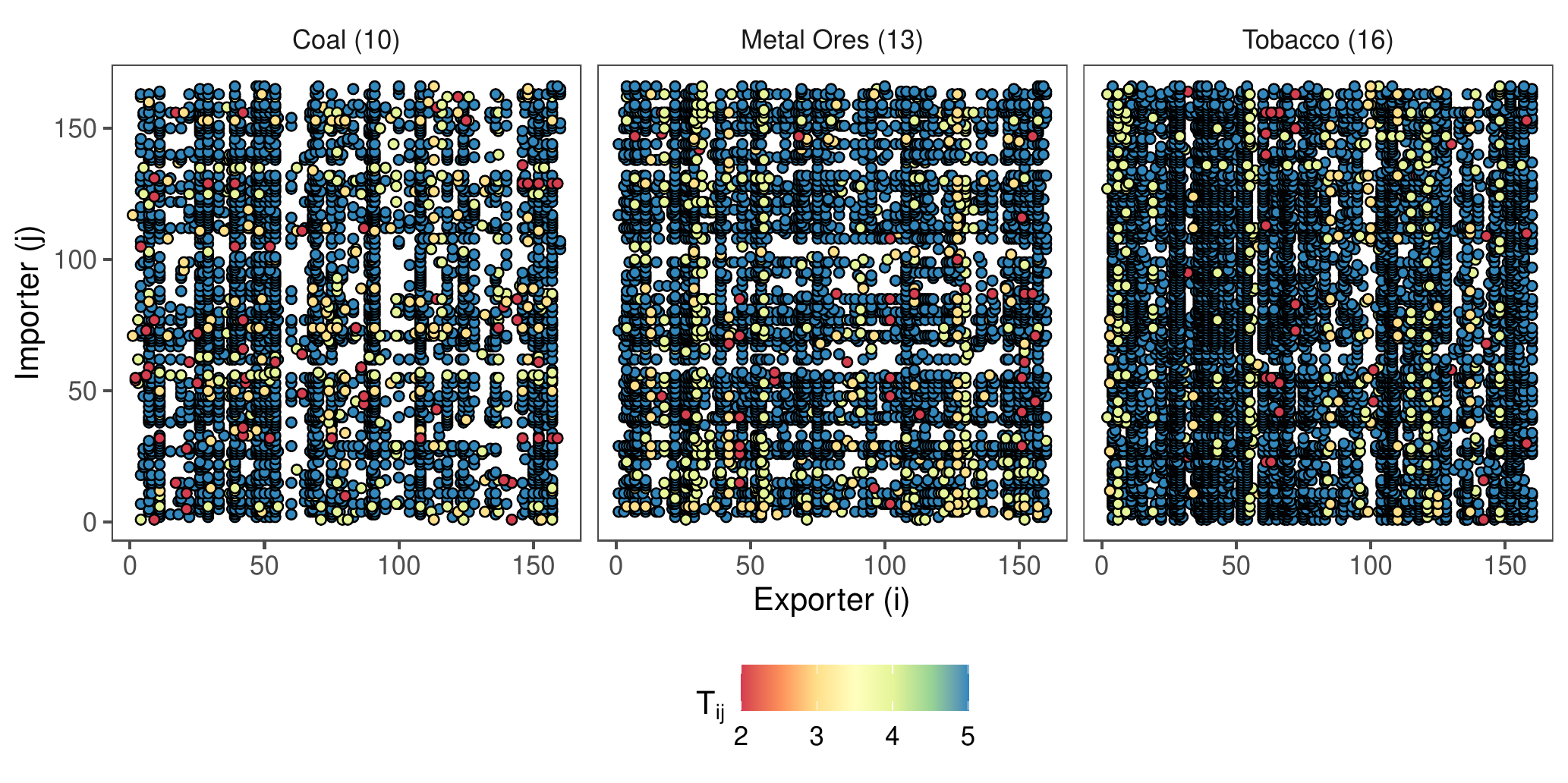}
\begin{minipage}{0.9\textwidth}
	\footnotesize
	\textbf{Notes:} Data taken from the replication package of WZ. Figure shows patterns of latent unbalancedness in the three most affected industries. Each point indicates a trading country pair. The points differ in color to show how many times we observe the corresponding country pair.
\end{minipage}
\label{fig:patterns}
\end{figure}

\clearpage

\section{Additional Simulation Results}

\subsection{Simulation Evidence -- More Detailed Results}

\begin{table}[!htbp]
\caption{Simulation Results -- $N = 200$, $T = 5$, and $\psi \in \{0, 0.1, \ldots, 0.8, 0.9\}$}
\centering
\begin{threeparttable}
\begin{tabular}{ccccccccc}
    \toprule
    $\psi$ & $\overline{N}$ & $n$ & \multicolumn{3}{c}{Bias (in \%)} & \multicolumn{3}{c}{Bias / SD} \\ \cmidrule(lr){4-6} \cmidrule(lr){7-9}
    & & & FE-PPML & ABC & SPJ & FE-PPML & ABC & SPJ \\ 
    \midrule
    0.0 & 200 & 200{,}000 & 0.371 & 0.016 & 0.005 & 0.677 & 0.029 & 0.010 \\ 
    0.1 & 180 & 180{,}005 & 0.385 & 0.013 & 0.001 & 0.684 & 0.022 & 0.001 \\ 
    0.2 & 160 & 160{,}005 & 0.420 & 0.021 & 0.005 & 0.725 & 0.036 & 0.009 \\ 
    0.3 & 140 & 140{,}000 & 0.444 & 0.010 & 0.008 & 0.741 & 0.017 & 0.013 \\ 
    0.4 & 120 & 120{,}005 & 0.493 & 0.013 & 0.012 & 0.774 & 0.020 & 0.018 \\ 
    0.5 & 100 & 100{,}005 & 0.578 & 0.034 & 0.000 & 0.836 & 0.048 & 0.000 \\ 
    0.6 & 80 & 80{,}000 & 0.677 & 0.040 & 0.011 & 0.902 & 0.052 & 0.015 \\ 
    0.7 & 60 & 60{,}000 & 0.848 & 0.063 & 0.022 & 1.008 & 0.074 & 0.025 \\ 
    0.8 & 40 & 40{,}005 & 1.187 & 0.140 & 0.034 & 1.195 & 0.138 & 0.033 \\ 
    0.9 & 20 & 20{,}005 & 2.177 & 0.512 & 0.094 & 1.589 & 0.366 & 0.062 \\ 
    \bottomrule
\end{tabular}
\begin{tablenotes}
\footnotesize
\item \textbf{Notes:} Table shows relative biases (in \%) and ratios of bias to standard deviation for different estimators for $\beta$: uncorrected FE-PPML, analytically corrected ABC, and split-panel jackknife corrected SPJ. Results are based on $10{,}000$ simulated samples for each $\psi$.
\end{tablenotes}
\end{threeparttable}
\end{table}

\subsection{Simulation Evidence -- Confirm Validity of Heuristic}

We conduct a simulation study to confirm the validity of the heuristic proposed by WZ. We first generate balanced panels with $N$ countries trading with each other for $T = 5$ periods, i.e. each simulated sample has initially $n = N^{2}T$ observations. Then $\psi N^2$ country pairs are dropped randomly so that the final sample has $n^{\ast} = (1 - \psi) N^2 T \approx N\overline{N}T$ observations. We consider different fractions of missing observations $\psi \in \{0, 0.1, \ldots, 0.8, 0.9\}$ and choose $N = \lfloor 20 / (1 - \psi) \rfloor$ so that $\overline{N} \approx 20$ remains roughly constant over $\psi$. We generate balanced panels following DGP I of WZ,
\begin{align}
    y_{ijt} &= \lambda_{ijt} \, \omega_{ijt} \, , \\
    x_{ijt} &= 0.5 \, x_{ijt-1} + \alpha_{it} + \gamma_{jt} + \eta_{ij} + \nu_{ijt} \, , \nonumber \\
    \omega_{ijt} &= \exp(- 0.5 \, \log(1 + \lambda_{ijt}^{- 2}) + (\log(1 + \lambda_{ijt}^{- 2}))^{- 0.5} \, z_{ijt}) \, , \nonumber \\
    z_{ijt} &= 0.3 \, z_{ijt-1} + \xi_{ijt} \, , \nonumber
\end{align}
where $i, j \in \{1, \ldots, N\}$, $t \in \{1, \ldots, T\}$, $\lambda_{ijt} = \exp(\beta \, x_{ijt} + \alpha_{it} + \gamma_{jt} + \eta_{ij})$, $\alpha_{it}$, $\gamma_{jt}$, and $\eta_{ij}$ are $\sim \iid \N(0, 0.25)$, $\nu_{ijt} \sim \iid \N(0, \sqrt{0.5})$, $x_{ij0} = \eta_{ij} + \nu_{ij0}$, $z_{ij0} \sim \iid \N(0, 1)$, $\xi_{ijt} \sim \iid \N(0, 0.91)$, and $\N(a, b)$ denotes the normal distribution with mean $a$ and variance $b$. We set $\beta = 1$ and consider the same estimators for $\beta$ as WZ: an uncorrected estimator (FE-PPML), an analytically bias-corrected estimator (ABC), and a split-panel jackknife bias-corrected estimator (SPJ). Our results are based on $10{,}000$ simulated samples for each $\psi$.

\begin{figure}[!htbp]
\caption{Estimates for $\beta$ for Different $\psi$}
\centering
\includegraphics[width=0.9\textwidth]{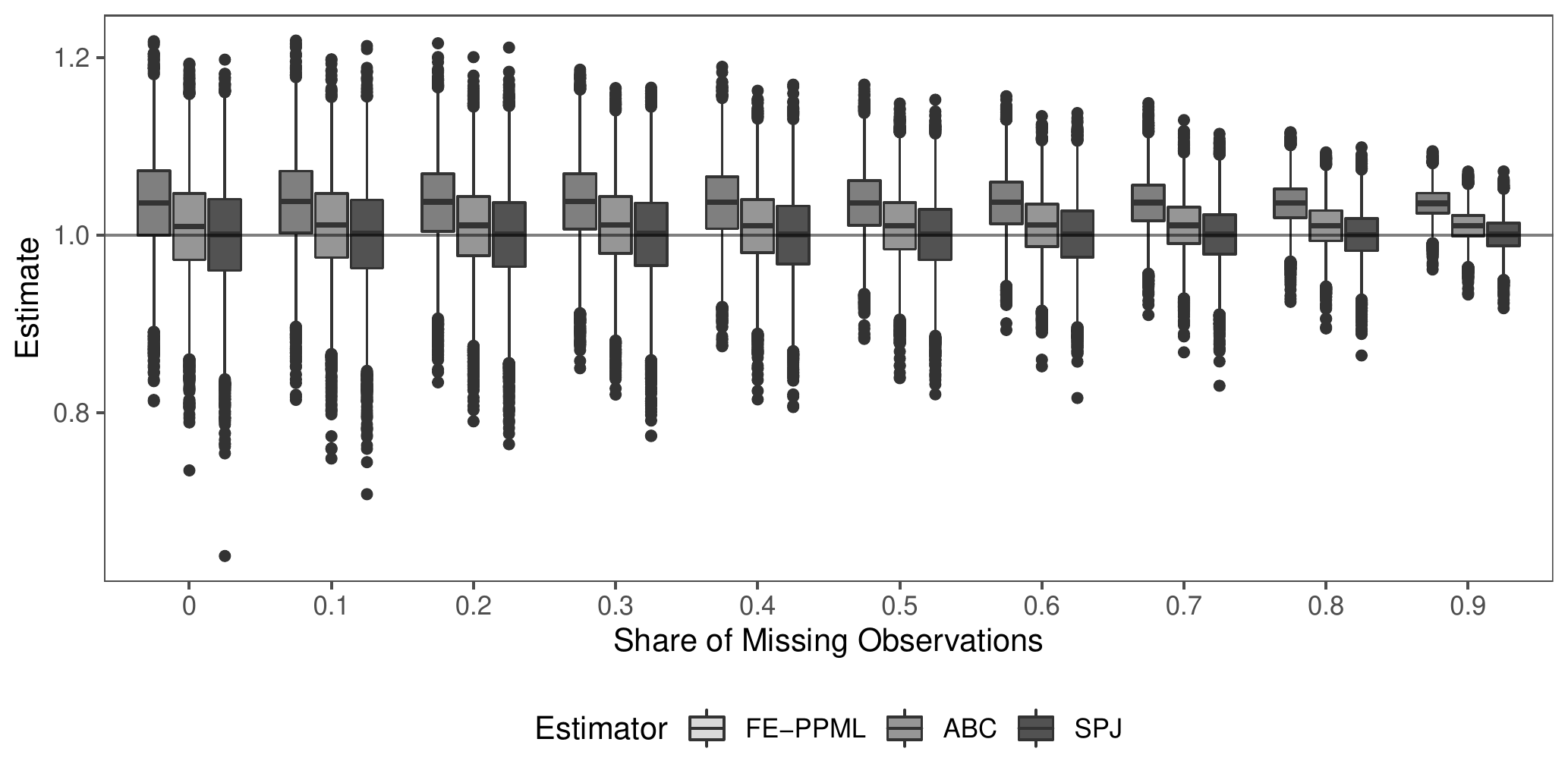}
\begin{minipage}{0.9\textwidth}
	\footnotesize
	\textbf{Notes:} Figure shows boxplots of estimates for $\beta$ from different estimators: uncorrected FE-PPML, analytically corrected ABC, and split-panel jackknife corrected SPJ. Results are based on $10{,}000$ simulated samples for each $\psi$.
\end{minipage}
\label{fig:results_nbar}
\end{figure}

Figure \ref{fig:results_nbar} summarizes the results. As predicted by the heuristic, we find that the biases remain constant over $\psi$, i. e. depend on the average number of trading countries ($\overline{N}$). Furthermore, we find that increasing $\psi$ decreases the dispersion of the estimators, as indicated by the boxes becoming narrower. This was to be expected since the sample size $n^{\ast}$ grows with $\psi$. However, this also means that inference gets worse as $\psi$ increases, since biases increase relative to the dispersion of the estimators. In other words, confidence intervals constructed around the estimators become narrower without any improvement in their centering by increasing the sample size.

\end{document}